\documentclass[twocolumn,showpacs,a4,prl]{revtex4} 
\usepackage{graphicx}%
\usepackage{dcolumn} 
\usepackage{bm} 
\makeatletter \def\btt#1{\texttt{\@backslashchar#1}}%
\DeclareRobustCommand\bblash{\btt{\@backslashchar}}%
\makeatother  
\begin{document}  
\preprint{paper.tex}  
\title{Nuclear Magnetic Resonance as  a probe of nanometre-size orbital textures in 
magnetic transition metal oxides}
\author{G. Papavassiliou$^1$, M. Pissas$^1$, M. Belesi$^2$, M. Fardis$^1$, 
D. Stamopoulos$^1$, A. Kontos$^1$, M. Hennion$^3$, J. Dolinsek$^4$, J. P. Ansermet$^{5,\star }$, 
and C. Dimitropoulos$^{1,\star }$} 
\affiliation{$^1$Institute of Materials Science, NCSR,  Demokritos, 
153 10 Aghia Paraskevi, Athens, Greece\\
$^2$Dept. of Physics, Ames Laboratory, Iowa State University, Ames, Iowa 50011\\
$^3$Laboratoire Leon Brillouin, CEA-CNRS, CE-Saclay, 91191 Gif sur Yvette, France\\
$^4$Josef Stefan Institute, Jamova 39, 61111 Ljubljana, Slovenia\\
$^5$Dept. of Physics, University of Illinois, Urbana, Illinois 1801, USA}
\altaffiliation{$^\star $Institut de Physique Exprimentale, EPFL-PH-Ecublens, 
1015-Lausanne, Switzerland} 
\date{\today }   
\begin{abstract} 
The study of strong electron correlations in transition metal oxides with 
modern microscopy and diffraction techniques unveiled a  fascinating  world of nanosize 
textures in the spin, charge, and crystal structure. Examples range from high $T_c$ 
superconducting cuprates and nickelates, to hole doped manganites and cobaltites. However, 
in many cases the appearance of these textures is accompanied with "glassiness" and 
multiscale/multiphase effects, which complicate significantly their experimental verification. 
Here, we demonstrate how nuclear magnetic resonance may be uniquely used to probe nanosize 
orbital textures in magnetic transition metal oxides. As a convincing example we show for 
the first time the detection of nanoscale orbital phase separation  in the ground state of 
the ferromagnetic insulator La$_{0.875}$Sr$_{0.125}$MnO$_3$.

\end{abstract} 
\maketitle  

It is widely accepted by now that orbital ordering is a key property, which controls the 
electronic behaviour of many transition metal oxides \cite{Tokura00,Saitoh01,Dagotto01,Hotta04}. 
A seminal case is LaMnO$_3$ \cite{Carvajal98,Murakami98}, where at temperatures lower than 
$T_{JT}\approx 780$K, cooperative Jahn-Teller (JT) distortions force an antiferro-orbital 
ordering within the {\it ab} plane. This orbital arrangement defines the magnetic order of 
the ground state  via the Goodenough-Kanamori-Anderson rules: Spins are coupled ferromagnetically 
in the {\it ab} planes and antiferromagnetically along the {\it c} axis, giving rise to the 
so called A-type antiferromagnetic (AFM) spin ordering. 
\begin{figure}[tbp] 
\centerline{\includegraphics[angle=0,width=7cm]{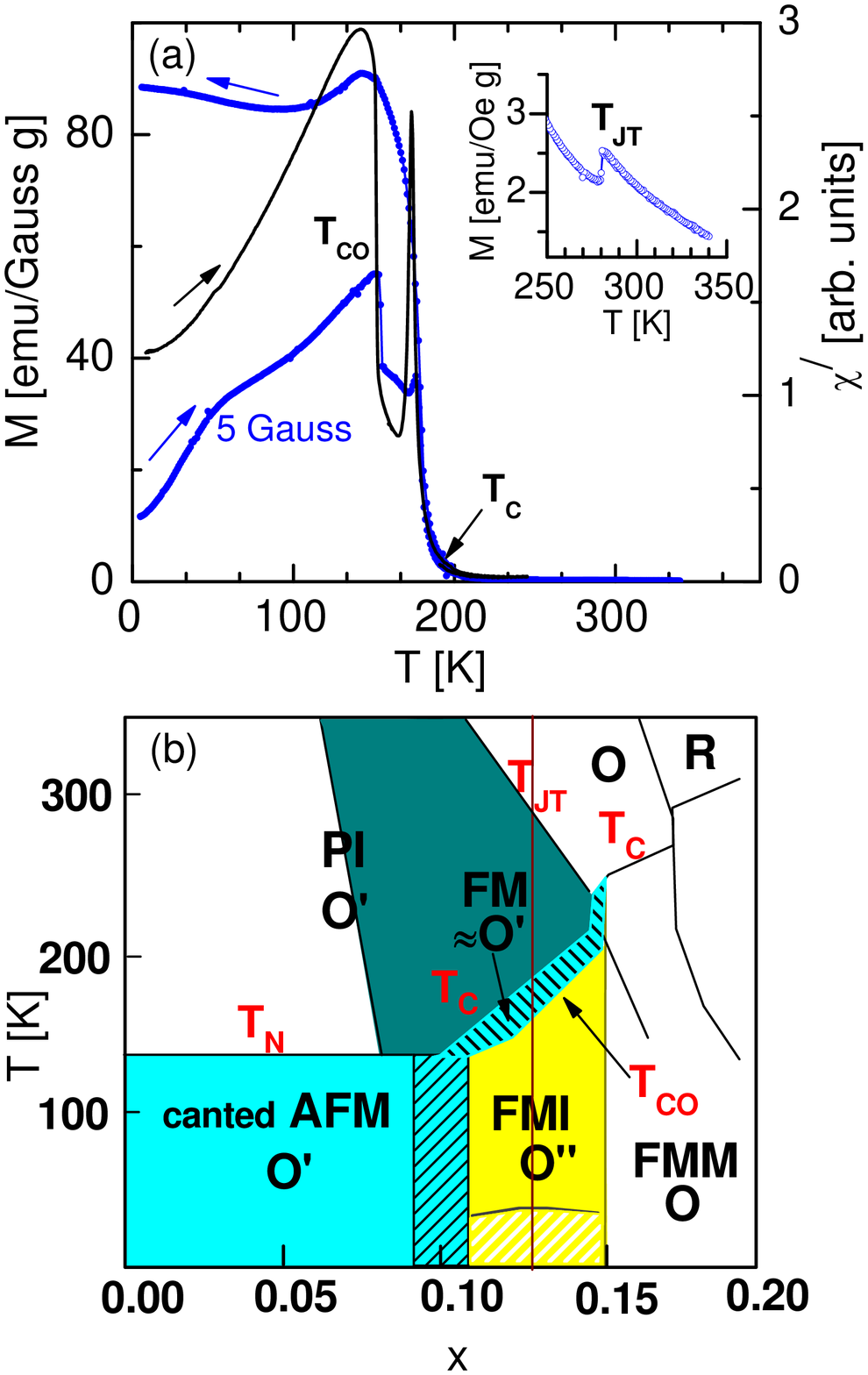}}
\caption{The magnetic and structural phase transitions of La$_{1-x}$Sr$_x$MnO$_3$ in the $T-x$ 
region of inerest. (a) SQUID magnetization measurements at $5$ Gauss (blue curve), and 
ac-susceptibility measurements (black curve), performed on a high quality $x=0.125$ single 
crystal. (b) The $T-x$ phase diagram, in the doping range $0\leq x\leq 0.2$, as reconstructed 
from refs. \cite{Paraskevopoulos00,Liu01}, by including magnetization and NMR data from the 
present work. The yellow-white hatched low temperature region arround $x=0.125$ is characterized 
by nucleation of FM O-type nanodomains into the O' matrix. (FMI=FM insulating, FMM=FM metallic).}
\label{Fig1} 
\end{figure}
By substituting La with a divalent cation such as Sr, the JT-active Mn$^{3+}$ sites are replaced 
by Jahn-Teller-inactive Mn$^{4+}$ sites, thus introducing into the system holes and energizing 
ferromagnetic (FM) coupling through the double exchange mechanism \cite{deGennes60}. New 
orbital structures are thus established \cite{Yamada96,Deisenhofer03,Geck04,Mizokawa00}. 

From the crystallographic perspective, by increasing doping $x$ the ground state of the 
La$_{1-x}$Sr$_x$MnO$_3$ (LSMO) changes from the JT distorted orthorombic O' phase (canted 
AFM insulating), in the low doping regime, to the - almost cubic - orthorombic O phase 
(FM and metallic) at $x\approx 0.20$, as shown in the lower pannel of Figure \ref{Fig1} 
\cite{Paraskevopoulos00,Liu01}. 
Around $x\approx 0.125$ the system exhibits a peculiar FM and charge ordered (CO) 
ground state with nominally "pseudocubic" crystal structure, which is known in the 
literature as O'' structure. The phase transition route for the $x=0.125$ system as a 
function of temperature is clearly seen in the SQUID and ac-susceptibility measurements, 
in the upper pannel of Figure \ref{Fig1}. Three successive phase transitions are observed 
at temperatures $T_{JT}\approx 283$K, $T_c\approx 183$K, and $T_{CO}\approx 150$K. The 
former transition (inset in the upper pannel of Figure \ref{Fig1}) is related with the 
passage from the high temperature pseudocubic O phase to the JT-distorted O' phase 
\cite{Geck04,Uhlenbruck99}, while the latter marks the transition to the CO and orbitally 
reordered \cite{Geck04} O'' phase. Besides, the large difference in the zero-field cooling 
and field cooling routes in the SQUID magnetization measurements is indicative of glassy 
freezing and metastability effects. Similar freezing effects have been observed in other 
Ca-doped manganites \cite{Papavassiliou01,Markovich04}, as well as in high $T_c$ 
superconducting cuprates \cite{Julien01}. Most spectacularly, upon x-ray irradiation 
the charge ordered phase  is partially destroyed below $40$K, being restored only after 
heating above $T_{CO}$ and subsequent cooling \cite{Kiryukhin99,Casa01}. The thermal 
history dependence of this effect and the observed photoinduced structural relaxation 
\cite{Kiryukhin99}, indicates that the O'' phase is extremely sensitive to the JT-induced 
strain fields, while showing a tendency to nanoscale phase separation \cite{Kiryukhin99}. 
Here, by using the NMR radiofrequency (rf) enhancement method \cite{Gossard59}, we 
uncover an important - but invisible until now - orbital phase separation in the ground 
state of $x=0.125$ LSMO. Specifically, we show that below $30$K the FM and charge 
ordered O'' matrix phase becomes unstable against the formation of FM nanodomains  
with the O-type orbital and crystal structure. 

To show this intriguing orbital nanotexturing, $^{139}$La NMR rf enhancement measurements 
in zero external magnetic field were performed on three high quality LSMO single crystals, 
with doping $x=0.075, 0.125$, and $0.20$. NMR in magnetic materials differs from 
conventional NMR techniques in several aspects \cite{Gossard59}. The most obvious 
difference is the presence of a spontaneous magnetic hyperfine field, 
$B_{hf}=\frac 1{\gamma \hbar }A\left\langle S\right\rangle$ at the position of 
the resonating nuclei, where $A$ and $\left\langle S\right\rangle $ are the hyperfine 
coupling constant and the average electronic spin, respectively. According to this formula, 
the hyperfine field $B_{hf}(La)$ at the position of the spinless ($S=0$) La sites in 
LSMO compounds is expected to reflect the average spin state of the surrounding Mn octant, 
as well as possible deformations of the Mn-O-Mn bonding, which alter the hyperfine 
coupling constant $A$ \cite{Papavassiliou01,Papavassiliou03}. 
A less apparent but very important difference is the existence of the rf enhancement 
\cite{Gossard59,Papavassiliou03}. In FM materials, very strong NMR signals are produced 
at extremely low rf irradiation fields $B_1$, due to coupling of the rf field with the 
magnetic moments of the electrons. Specifically, the applied transverse rf field $B_1$ 
produces an internal field $B_1/(1+N\chi)$, where 
$N$ is the demagnetizing factor, and $\chi$ the magnetic susceptibility. The magnetization 
then rotates through an angle $B_1/(1+N\chi )B_A$, where $B_A$ is the magnetic 
anisotropy field. The rotation of the magnetization at the radiofrequency thus produces 
an effective rf field $B_{1,eff}$ sufficiently stronger than the applied $B_1$ by the 
factor $n=[B_{hf}/(1+N\chi )B_A]$. We notice that the rf enhancement factor $n$ 
would decrease in an external magnetic field, while at low fields both domain rotations 
and domain-wall displacements are expected to contribute to the transverse magnetization. 
However, in strongly inhomogeneous systems like CMR manganites, rf enhancement experiments 
reflect directly the local $B_A$ in the various magnetic subphases \cite{Papavassiliou03}.

\begin{figure}[tbp] 
\centerline{\includegraphics[angle=0,width=8cm]{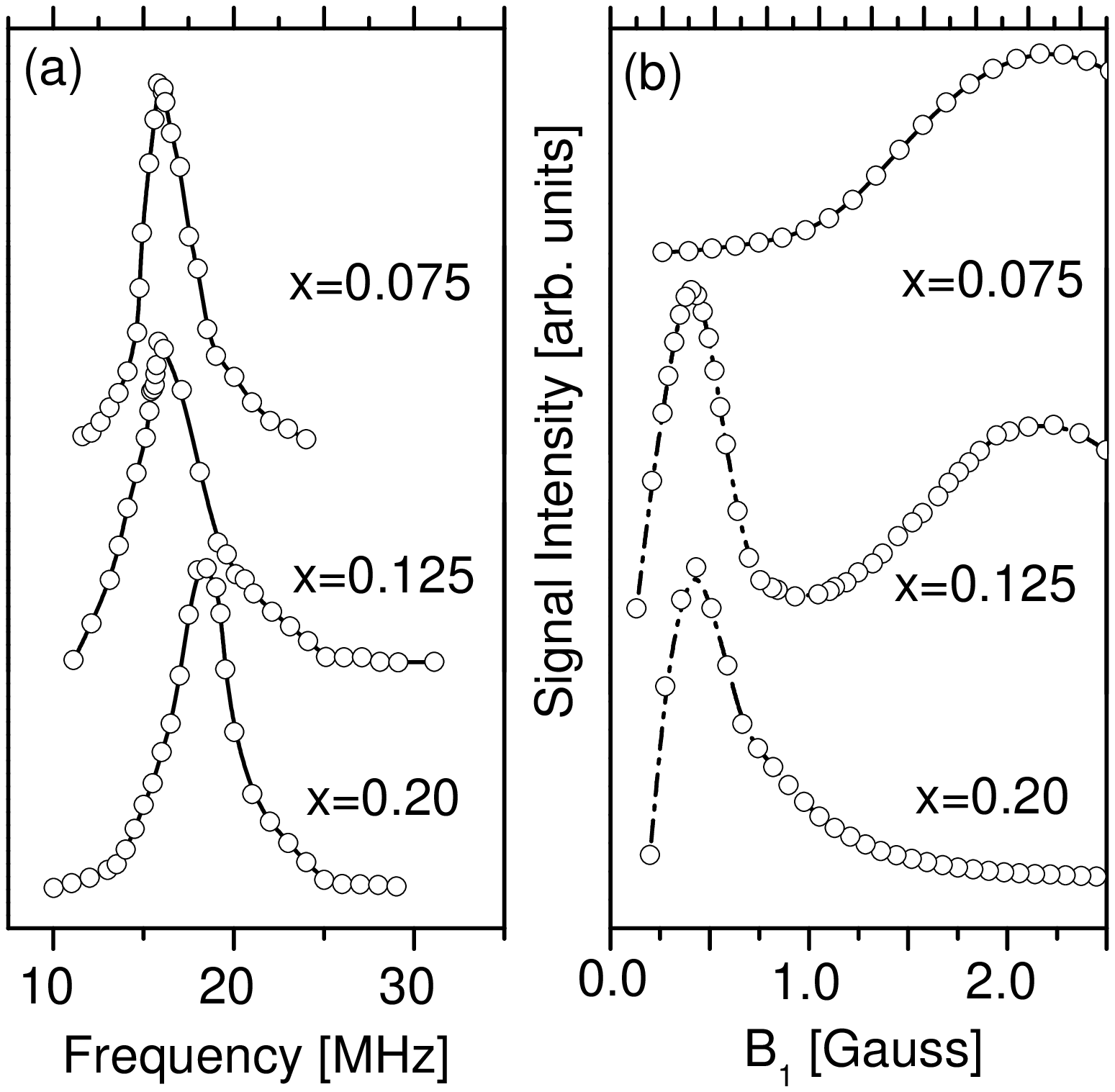}}
\caption{(a) Zero field $^{139}$La NMR spectra of LSMO for $x=0.075, 0.125$, and $0.20$, 
at $5K$. (b) The $^{139}$La NMR signal intensity $I$ vs. the rf field $B_1$ for $x=0.075, 
0.125$, and $0.20$.}
\label{Fig2} 
\end{figure}

Figure \ref{Fig2} shows zero field $^{139}$La NMR line shape measurements and rf 
enhancement plots at $5$K for the three systems under investigation. The line shapes 
were acquired by applying a two pulse spin-echo technique with pulse widths 
$t_{p1}=t_{p2}=0.6 \mu $ sec,  after recording the integrated spin-echo signal 
intensity $I$ at successive irradiation frequencies. The rf enhancement experiments 
were performed by recording $I$ as a function of the applied rf field $B_1$. The 
obtained $I$ vs. $B_1$ curves follow an asymmetric bell-shaped law with maximum at 
$n\gamma B_1\tau =2\pi /3$, where $\tau $ is the rf pulse duration, $\gamma $ the 
nuclear gyromagnetic ratio, and $n$ the rf enhancement factor \cite{Papavassiliou97}. 
According to the measurements no major changes are observed in the line shapes of 
the investigated samples, either with doping $x$ (Figure \ref{Fig2}a), or with 
temperature variation (not shown in the plot). Specifically, by increasing $x$ the 
NMR spectra shift only slightly in frequency from $\approx 16$ MHz for $x=0.075$ 
to $\approx 18$ MHz for $x=0.20$, indicating a nearly FM environment for the 
resonating nuclei in all three cases. The observation of FM NMR signals for 
$x\geq 0.075$ is in agreement with "in field" $^{139}$La NMR line shape measurements 
for LSMO, with $0\leq x\leq 0.15$, which show a very sharp changeover from the 
AFM to the FM phase at $x\approx 0.05$ \cite{Kumagai99}. Besides, neutron scattering 
experiments have shown the presence of a single "FM modulated" canted AFM state 
for $0.06\leq x\leq 0.1$ \cite{Hennion00,Kober04}, comprised of FM plateletes, 
a few unit cells large on the {\it ab} plane, which are exchange coupled 
through the canted AFM matrix. 

\begin{figure}[tbp] 
\centerline{\includegraphics[angle=0,width=7cm]{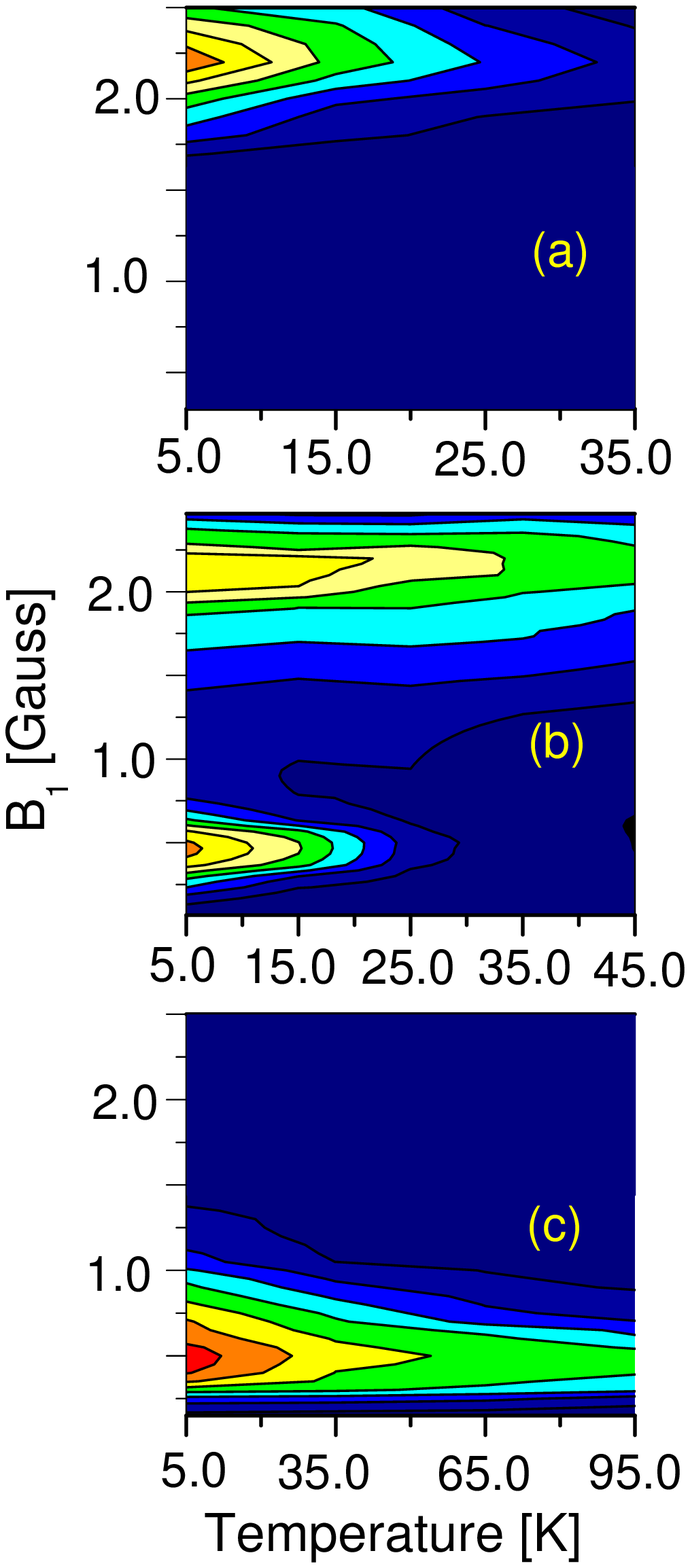}}
\caption{$^{139}$La NMR rf enhancement experiments for La$_{1-x}$Sr$_x$MnO$_3$ with 
(a) $x=0.075$, (b) $x=0.125$, and (c) $x=0.20$. The contour plots show the NMR signal 
intensity $I$ as a function of $B_1$ and $T$. There is a significant difference in 
the rf enhancement by varying doping. The plot in the middle shows clearly that the 
ground state of the LSMO $x=0.125$ is a mixture of two phases differing in their 
magnetocrystalline anisotropy.}
\label{Fig3} 
\end{figure} 

Contrary to the line shape measurements, a significant difference as a function of 
doping and temperature is observed in the rf-enhancement experiments. According to 
Figure \ref{Fig2}b the value $B_{1,max}$, where the maximum NMR signal is obtained, 
is shifted from $2.2$ Gauss for $x=0.075$ (with the JT-distorted O' type crystal 
structure) to $0.5$ Gauss for $x=0.20$ (with the orthorombic, and nearly isotropic 
O crystal structure). Most important, for $x=0.125$ the rf enhancement curve at 
$5$K is a superposition of the corresponding curves for the $x=0.075$ and $0.20$ 
systems, indicating the spontaneous spliting of the system in two phases with 
indistinguishable spin-structures, but different magnetocrystalline anisotropy. 
The evolution of this exciting phase separation by varying temperature can be 
nicely followed in the contour plots of Figure \ref{Fig3}. For $T>30$K a single 
phase is observed in Figure \ref{Fig3}b, which can be attributed to the O'' 
structure with antiferro-orbital ordering \cite{Endoh99}, while at $T\leq 30$K 
a second phase component appears, resembling the O structure of the $x=0.20$ 
system. We stress that synchrotron X-rays diffraction experiments observe only 
O''-type nanodomains with average size $\approx 30-35$ nm \cite{Kiryukhin99}. 
Evidently, the O-type phase component is confined into smaller "invisible" 
regions, which increase in size and become detectable only after strong 
illumination with X-rays \cite{Kiryukhin99}. The nucleation of such O-type 
islands explains the fast decrease of the antiferro-orbital ordering, observed 
with resonant X-ray scattering experiments in the low temperature regime of 
the O'' phase \cite{Endoh99}. It is also worth to notice the similarity in 
the contour plots of the O' and O" phases. A possible explanation is that at 
low doping, such as $x=0.075$, the observed NMR signals are solely produced 
in FM plateletes \cite{Hennion00,Kober04} which are precursors of the O'' 
phase. However, the size ($\approx 2$ nm) and the strong coupling of such 
plateletes with the canted AFM matrix state \cite{Hennion00,Kober04}, suggests 
that their anisotropy should be rather determined by the anisotropy of the 
O' matrix state. It is thus possible that despite the different orbital 
hybridization of the O' and O'' phases, the similarity in their orbital 
ordering (i.e. antiferro-orbital ordering for both phases \cite{Endoh99}) 
gives rise to comparable $B_A$ values. 

\begin{figure}[tbp] 
\centerline{\includegraphics[angle=0,width=8cm]{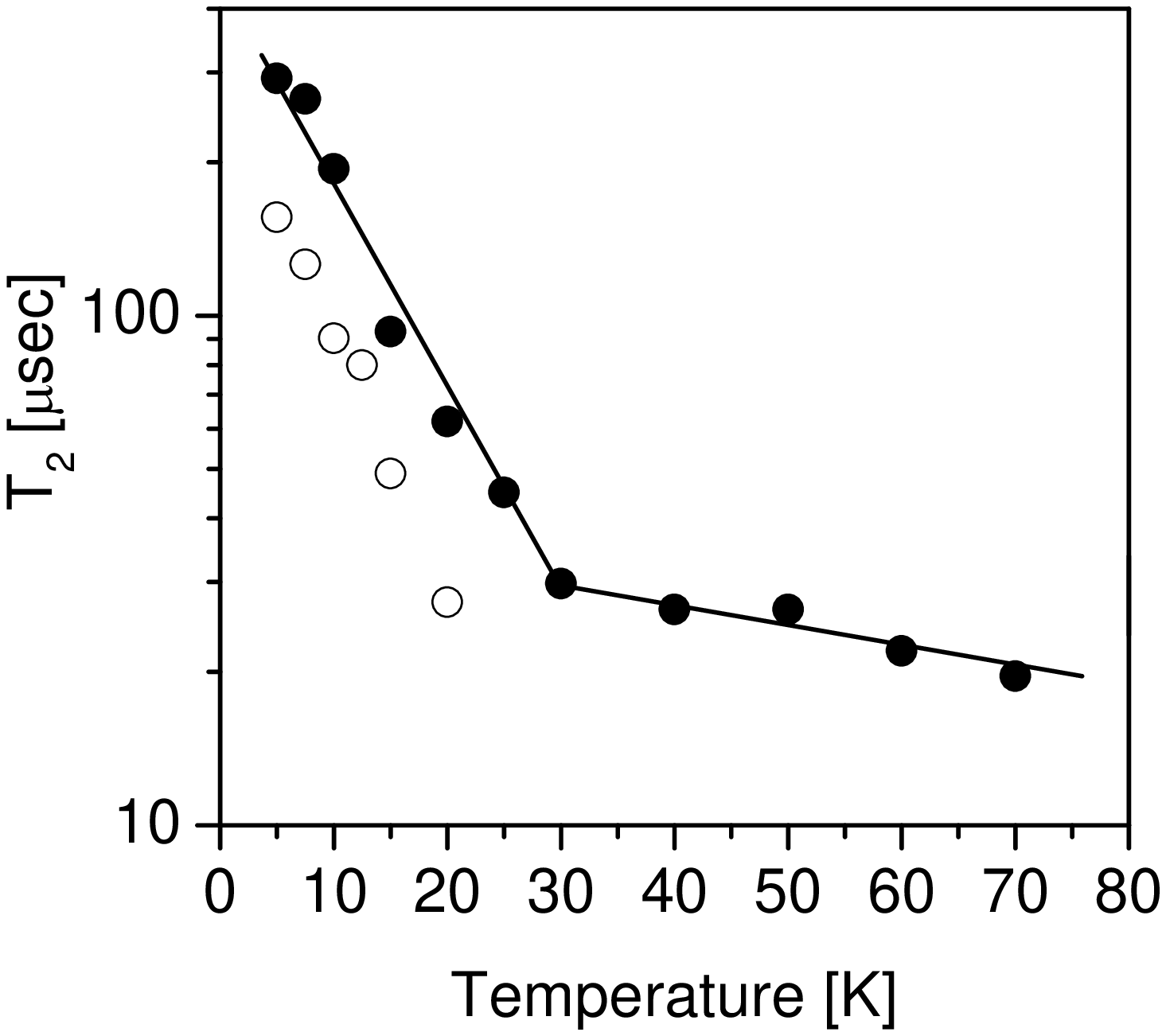}}
\caption{$^{139}$La NMR spin-spin relaxation time $T_2$ of La$_{0.875}$Sr$_{0.125}$MnO$_3$ 
as a function of temperature. Measurements were performed at two different rf fields, 
i.e. at $B_1=0.5$ Gauss (open circles) and $2.2$ Gauss (filled circles), which 
correspond to signals from the O and O'' structures, respectively. It is observed 
that the appearance of O-type domains is accompanied with a slope change in 
the $T_2$ of the signal from the O'' domains.}
\label{Fig4} 
\end{figure}  

Further information about this orbital phase separation, has been obtained by 
performing $^{139}$La NMR spin-spin relaxation time ($T_2$) measurements on 
both phase components. $T_2$s were measured with the two pulse spin-echo technique, 
as previously described, by varying the time interval between the two pulses and 
recording the decay of the spin-echo signal intensity. The experimental data in 
Figure \ref{Fig4} show that the appearance of the O phase component is accompanied 
with a slope change in the $T_2$ vs. $T$ curve of the 
O'' phase component. Most important, below $30$K the $T_2$ curves for both phase 
components exhibit the same slope, which is indicative of a similar evolution in 
their spin dynamics by cooling. Hence, the picture that emerges from Figures \ref{Fig3} 
and \ref{Fig4} is that the O'' phase is metastable, whereas below $30$K droplets 
of the O phase at nanometer-scale start to nucleate. A possible explanation for 
this behaviour is that strain fields, characterizing the JT-distorted O' phase, 
are decreasing gradually in the O'' phase, giving rise to undistorted FM metallic 
islands below a certain transition temperature. A new region, is thus defined 
in the $T-x$ phase diagram, which is shown as yellow-white hatched region in 
Figure \ref{Fig1}. 

In summary, the direct relation between $B_A$ and the NMR rf enhancement in 
low doped LSMO, allowed us to unveil an unexpected orbital nanophase separation 
in the ground state of LSMO, $x=0.125$. This kind of phase separation appears to 
underlie to the photoinduced phase segregation in LSMO \cite{Kiryukhin99}, 
as well as in other CMR manganites \cite{Miyano97}. In general, the method we 
have employed here is applicable in many other magnetic transition metal oxides, 
where orbital rearangements are not reflected on the spin structure, but give 
appreciable differences in the local magnetocrystalline anisotropy.

\end{document}